\documentclass{aa}
\usepackage{graphicx}
\usepackage{txfonts}
\usepackage{natbib}
\usepackage{subfigure}
\begin{document}

%   \thesaurus{10494}      

   \title{Bending instabilities at the origin of persistent warps: a new constraint on dark matter halos}

   \author{Y. Revaz \and D. Pfenniger}

   \offprints{Y. Revaz}

   \institute{Geneva Observatory, University of Geneva, CH-1290 Sauverny, Switzerland\\
              email: Yves.Revaz@obs.unige.ch}

   \date{Received -- -- 20--/ Accepted -- -- 20--}

   \abstract{ A substantial fraction of the warps in spiral galaxies
     may result from bending instabilities if the disks are
     essentially self-gravitating.  With N-body simulations, we show
     that galaxies with self-gravitating disks as thick as HI disks
     are subject to bending instabilities generating 
     S-shaped, U-shaped or asymmetric warps.  
     S-shaped warps persist during several rotations
     and keep the line of node straight.  The warp amplitudes
     generated by bending instabilities remain however modest.
     Other factors must be invoked for extreme warped disks. However,
     bending instabilities can account for most of the cases reported
     in optical surveys, where the warp angle is 
     generally less than $5^\circ$.
     This mode of warping is very sensitive to the disk flattening.
     It also constrains the fraction of dark matter 
     distributed in the disk and in the dark halo.
   
   }

   \maketitle

%
%________________________________________________________________

\section{Introduction}

Since the first observation of the Milky Way warp at the end of the
fifties \citep{kerr56,burke57}, half a century has passed without a clear
explanation of this feature \citep[see][for a review]{binney92}.  The
warp problematic comes essentially from its high frequency in spirals.
In the local group, including the Milky Way, the three dominant spirals
are warped \citep{roberts66,rogstad76,newton77} and among nearby
galaxies they are common \citep{sancisi76,bosma81b}.  Statistics of
warps in HI \citep{bosma91,richter94,garciaruiz98} and in the optical
band \citep{reshetnikov98,reshetnikov99,sanchez90,sanchez03} reveal
that more than half the spiral galaxies are warped and asymmetric.
This implies that warps are either frequently or continuously
generated, or persistent over several dynamical times. Disks are
easily warped when subject to an external torque, generated for
example by gravitational forces during interactions
\citep{quinn91,hernquist91,huang97} or accretion events
\citep{jiang99,revaz01b,lopez02}, or by magnetic forces in the gas
\citep{battaner90}.  The challenge is to understand how isolated
galaxies (in appearance) can conserve their warp with a relatively
straight line of nodes (LON), avoiding the winding problem early
noticed by \citet{kahn59}.  Solutions like normal modes of vibration
(bending waves) have been proposed \citep{lynden-bell65} .
Unfortunately \citet{hunter69} showed that no discrete modes exist in
disks with soft edges. If, in the presence of a dark halo, discrete
modes can exist \citep{sparke84,sparke88} they will be quickly damped
by dynamical friction \citep{nelson95,dubinski95}.

However, as pointed out by \citet{sellwood96}, previous studies have
treated the case of a cold
razor-thin disk. When taking
into account the velocity dispersions necessary to keep the disks
stable against axisymmetric instability, disks may be subject to a
bending instability (also named fire-hose instability).  Bending
instabilities were first studied by \citet{toomre66} in an
infinite slab of finite thickness.  The instability occurs when the
centrifugal force of particles following a bend in the slab is higher
than the gravitational restoring force of the slab itself.
A formal description of the instability can be found in \citet[p.37]{fridman84}.
\citet{araki85} showed that an infinite slab is stable when the ratio
of the vertical velocity dispersion $\sigma_z$ to the velocity
dispersion in the plane $\sigma_{u}$ exceeds $0.293$.
\citet{sellwood94} and \citet{merritt94} have studied this instability
in realistic disks of counter-rotating stellar components. They found
discrete bending modes $m=0$ and $m=1$ that could persist in disks
even without counter-rotation.  \citet{sellwood96} studied the growth
of such modes in axisymmetric cases and found that they are strongly
Landau damped \citep{fridman84}. He also mentioned that some warm disks support
long-lived axisymmetric flapping oscillations.

Since the end of the sixties, most of the works on the warp problem
have been built on the assumption of a disk embedded in a massive
non-rotating dark halo.  However, there are now many observations
that let us think that a substantial fraction of dark matter may
be present in outer galactic disks, in the form of rotationally supported
cold and clumpy gas \citep{pfenniger94a,pfenniger94b}.

The presence of a weak but existing star formation rate far from the
center of galaxies has been reported by different authors
\citep{naeslund97,smith00,cuillandre01}.  This star formation rate
reveals that molecular gas must be present in abundance in the outer
disk of galaxies.  Moreover, it has been pointed out long ago by
\citet{bosma78,bosma81a} that in samples of galaxies, the ratio
between the dark matter and HI surface density is roughly constant
well beyond the optical disk \citep[see
also][]{carignan90,broeils92,hoekstra01}. These observations suggest
that the dark matter follows the HI distribution along the disk.  More
recently, the study of the spiral pattern of NGC 2915 has suggested
that the very extended HI disk in this galaxy is supported by a quasi
self-gravitating disk \citep{bureau99,masset03}.

In the context of the warps, massive disks are interesting because the
kinetic pressure support is then lower than the rotational support and
such disks are in a regime where the Araki criterion \citep{araki85}
is relevant, thus bending instabilities are expected.

In the first part, we examine the evolution of bending instabilities in
different self-consistent galactic models, where the mass is dominated
by a heavy disk. The models differ essentially in the thickness of the
heavy disk.  We show that between the very unstable thin disks and the
stable thick disks, a regime exists where bending instabilities
spontaneously generate persistent warps.  In the second part, a
semi-analytic analysis allows us to fix constrains on the validity of
the previous results, when a dark halo is also taken into account.
The values of the mass and flattening of the dark halo compatible with
bending instabilities is determined.

%__________________________________________________________________

\section{The galaxy model}\label{model}

%__________________________________________________________________

Our galactic mass model is composed of a bulge, an exponential stellar
disk and a heavy disk made of HI gas and dark matter proportional to
it.
\begin{enumerate}
\item  
The bulge density profile is a flattened Plummer model:
\begin{equation}
  \rho_{\rm{b}}(R,z) \propto {\left( 1+ \frac{R^2}{a^2} + \frac{z^2}{b^2} \right)^{-5/2}},
   \label{bulge}
\end{equation}
where $a=1\,\rm{kpc}$ and $b=0.25\,\rm{kpc}$.

\item
The exponential stellar disk takes the usual form:
\begin{equation}
  \rho_{\rm{d}}(R,z) \propto e^{-R/H_R}\,e^{-|z|/H_z} ,
   \label{disk}
\end{equation}
where the radial and vertical scale length are
respectively $H_R=2.5$ and $H_z=0.25\,\rm{kpc}$.

\item 
To comply with the HI observations and slowly varying rotation
curves, the heavy disk density profile is:
\begin{equation}
        \rho_{\rm{hd}}(R,z) \propto \frac{e^{-{R_{\rm{hd}}}/{R}}}{R} \,
        \frac{e^{-{|z|}/{h_z(R)}}}{h_z(R)} \,
        \arccos\left({R}/{R_{\rm{hd}}^{\rm{max}}} \right).
        \label{noire_adopte_2}
\end{equation}
For $R>R_{\rm{hd}}=7\,\rm{kpc}$, the heavy disk surface density
decreases as $R^{-1}$. It is smoothly truncated by the arccosine
function at $R_{\rm{hd}}^{\rm{max}}=35\,\rm{kpc}$ in order to keep an
approximately constant rotation curve up to the disk edge.  In
agreement with observations showing that dark matter is not dominant
at the center of galaxies \citep{weiner01,bissantz03}, the heavy disk
surface density drops to zero at the center if $R_{\rm hd}>0$.  To be
consistent with the observed flaring of HI in the Milky Way
\citep{burton92}, the vertical scale height $h_z(R)$ is radius
dependent and is written as:
\begin{equation}
        h_z(R) = h_{z0}\, e^{{R}/{R_{\rm{f}}}},
        \label{hz_2}
\end{equation}
where $h_{z0}$ is the vertical scale height at the center and
$R_{\rm{f}}$ the flaring radial scale-length.  
\end{enumerate}

\begin{figure}
\resizebox{\hsize}{!}{\includegraphics[]{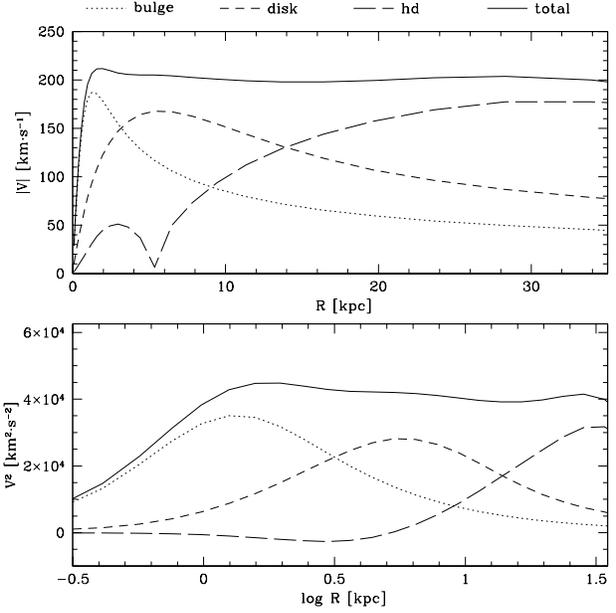}}
\caption{Contribution of each component to the total rotation curve. The upper panel shows a 
classical representation of the rotation curve. In the bottom panel, 
the square of the circular velocity is plotted as a function of 
$\textrm{log}R$. The latter representation enhances the effective contribution of
each component to the total rotation curve \citep{kalnajs99}.}
\label{rc}
\end{figure}

The respective masses
of the three components are $M_{\rm{b}}=0.15\cdot10^{11}\,\rm{M_\odot}$
for the bulge, $M_{\rm{d}}=0.46\cdot10^{11}\,\rm{M_\odot}$ for the exponential
stellar disk, and $M_{\rm{hd}}=1.61\cdot10^{11}\,\rm{M_\odot}$ for the heavy 
disk.  With these values the rotation curve is approximately flat up to
$R=35\,\textrm{kpc}$. Fig.~\ref{rc} shows the contribution of each component to the
total rotation curve. 
The lack of HI or dark matter at the center of the galaxy implies a negative 
circular velocity squared contribution by the heavy disk component 
(see the bottom panel of Fig.~\ref{rc}).
In the upper panel, where the norm of the circular velocity is traced as a function of radius, 
it produces a bump around $3\,\rm{kpc}$.

The initial vertical velocity dispersion $\sigma_z$ is found by
satisfying the equilibrium solution of the stellar hydrodynamic
equation in cylindrical coordinates \citep[see
e.g.,][p.\,199]{binney87} separately for each of the bulge, stellar
disk and heavy disk components,
\begin{equation}
        \rho_i \, {\sigma_z}_i^2 = \int_z^\infty\! dz \,\rho_i\, \partial_z \Phi         
        ,
        \label{st1}
\end{equation}
where the index $i$ can label any component.  These calculations are
carried out on the polar coordinate potential solver of the
Particle-Mesh $N$-body code described in \cite{pfenniger_friedli93}.
The ratio between the tangential velocity dispersion $\sigma_\phi$ and
the radial one $\sigma_R$ is derived from the epicycle approximation
\citep[][ p.\,125]{binney87}:
\begin{equation}
        \frac{\sigma_{\phi}^2}{\sigma_R^2}
        = \frac{\kappa^2}{4\Omega^2}
        = \frac{R^{-3}\partial_R\left(R^3\partial_R \Phi\right)_{z=0}} 
             {4R^{-1}\left(\partial_R \Phi\right)_{z=0}}
        ,
        \label{st2}
\end{equation}
where $\kappa$ is the radial epicyclic frequency, and $\Omega$ the
rotation frequency, determined from the potential $R$-derivatives as
indicated in Eq.~(\ref{st2}).  Since in this approximation the radial
oscillations are decoupled from the vertical oscillations, the ratio
$\sigma_z/\sigma_R$, which is precisely the Araki criterion, is not
constrained.  This allows us to choose the radial velocity dispersion.
We have used the following relation:
\begin{equation}
        \frac{\sigma_z^2}{\sigma_R^2}
        = \beta^2\,\frac{\kappa^2}{\nu^2}
        = \beta^2\,\frac{R^{-3}\partial_R\left(R^3 \partial_R \Phi\right)_{z=0}}
                {\left(\partial_z^2\Phi\right)_{z=0}}        ,
        \label{sr}
\end{equation}
where $\nu$ is the vertical epicycle frequency, determined from the
potential $z$-derivatives as indicated in Eq.~(\ref{sr}).  The factor
$\beta$ is constant of order 1 and chosen to fix the Savronov-Toomre
radial stability parameter $Q = \kappa\,\sigma_R/3.36\, G\, \Sigma$, where
$\Sigma = 2\int_0^\infty\!dz\, \rho(R,z)$ is the total surface density.

The found velocity dispersions are then used to distribute the model
particles in velocity space and to find the rotation velocity for each
component with the Jeans equations.

%The factor $\beta$ is
%sets to $1.3$ and insures that the Safronov-Toomre parameter $Q$
%ranges between $1$ and $2$. 
%
\begin{figure}
\resizebox{\hsize}{!}{\includegraphics[]{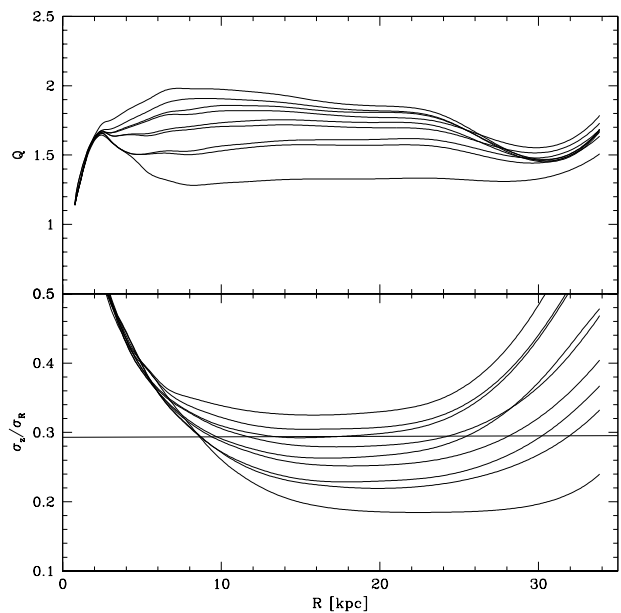}}
\caption{
  The radial stability parameter $Q$ (top) and the ratio
  $\sigma_z/\sigma_R$ (bottom) as a function of the galactic radius.
  In both graphics, the curves correspond, from bottom to top, in the
  interval $10$-$20$, to the models of increasing thickness, 01, 02,
  07, 03, 08, 04, 09, 05, 06, respectively.}
\label{szrq}
\end{figure}
%

%__________________________________________________________________

\section{The simulations}

%__________________________________________________________________

\subsection{Parameters}

In order to explore the existence and evolution of bending
instabilities, we have run a set of 9 models that differ in the
thickness of the heavy disk, parameterized by $R_{f}$ and $h_{z0}$.
Values are chosen such that the minimal ratio $\sigma_z/\sigma_R$
in the disk ranges in an interval around Araki's limit
\footnote{Araki's criterion has been derived in the case of an
  infinite slab.  However, it remains locally a good first order
  approximation when the rotation support exceeds the radial pressure
  support \citep{sellwood96}.}.
At equilibrium, $\sigma_z^2$ is proportional to the thickness of the
disk.  From Eq.~(\ref{sr}), the ratio $\sigma_z/\sigma_R$
strongly depends on the second $z$-derivative of the potential.  Thus,
increasing the thickness of the disk increases the ratio
$\sigma_z/\sigma_R$.  Moreover, since $\sigma_z$ and $\sigma_R$ are
linked by Eq.~(\ref{sr}) and since $\kappa$ and the surface density
$\Sigma$ are almost independent of the vertical thickness, thinner
disks are also more unstable with respect to the radial stability 
parameter $Q$.  In Fig.~\ref{szrq}, for each model, $Q$ and
$\sigma_z/\sigma_R$ are traced as functions of the radius.  With the
parameter $\beta$ fixed to $0.77$, all models are slightly sub-critical
with respect to the generation of spiral density waves.
\begin{table}
    \begin{tabular}{c | c c c c c c}
    \hline
    \hline
     Models     & 01 & 02 & 03 & 04  & 05 & 06 \\
    \hline
    \hline
    $R_{f} $ & 40 & 40 & 40 & 40 & 40 & 40 \\   
    $h_{z0}$ & 0.05 & 0.15 & 0.25& 0.35 & 0.45 & 0.55 \\  
    $\left(\sigma_z/\sigma_R\right)_{\rm{min}}$ & 0.186 & 0.218 & 0.248 & 0.272 & 0.295 & 0.314 \\
    $\zeta$ & 0.645 & 0.648 & 0.651 & 0.654 & 0.658 & 0.661\\  
    
    \hline
    \hline
    Models & 07 & 08 & 09\\
    \hline
    \hline    
    $R_{f} $ & 30 & 30 & 30 \\   
    $h_{z0}$ & 0.15 & 0.25 & 0.35\\
    $\left(\sigma_z/\sigma_R\right)_{\rm{min}}$ &  0.226 & 0.258 & 0.286 \\
    $\zeta$ & 0.648 & 0.652 & 0.656\\ 
    
    \hline
    \end{tabular}
    \caption[]{Parameters $R_{f} $ and $h_{z0}$ for the different models.
    The respective flattening $\zeta$ of the total potential is also indicated.}
    \label{params}
\end{table}

\begin{figure}
\resizebox{\hsize}{!}{\includegraphics[angle=-90]{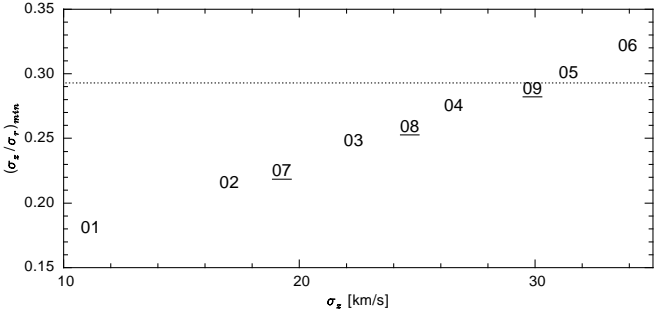}}
\caption{Ratio $\sigma_z/\sigma_R$ as a function of the vertical
  dispersion $\sigma_z$ at $R=15\,\rm{kpc}$.  The values are taken at
  the radius where $\sigma_z/\sigma_R$ is minimum.  The dotted
  line corresponds to Araki's limit.}
\label{szszr2}
\end{figure}

Fig.~\ref{szszr2} shows the distribution of the models in the plane 
$\sigma_z$--$\sigma_z/\sigma_R$, at the radius $R$ where $\sigma_z/\sigma_R$ is minimal.  
Thin disks are found at lower left end while thick disks are found at the upper right end.  
The dotted line corresponds to Araki's limit.
Parameters for each model, as well as $\left(\sigma_z/\sigma_R\right)_{\rm{min}}$ and 
the flattening $\zeta$ of the potential are summarized in Table~\ref{params}.  
The potential flattening is defined as the ratio of minor over major axis of the iso-potentials.

The models are evolved with the Barnes-Hut treecode parallelized to
run on GRAVITOR, the 132-processor Beowulf cluster of the Geneva
Observatory.  To trust thin disk transverse instabilities it is
important to not use an anisotropic Poisson solver such as a polar
grid mesh, therefore, despite being much more CPU time consuming, a
treecode approach is preferred.  Each simulation contains
$2^{18}=262\,144$ particles of equal mass.  The softening length is
0.15\,kpc, and the time-step is 0.5\,Myr.  If not indicated, the
length unit is the kpc, and the time unit is the Myr.

\begin{figure*}
\resizebox{\hsize}{!}{\includegraphics[angle=-90]{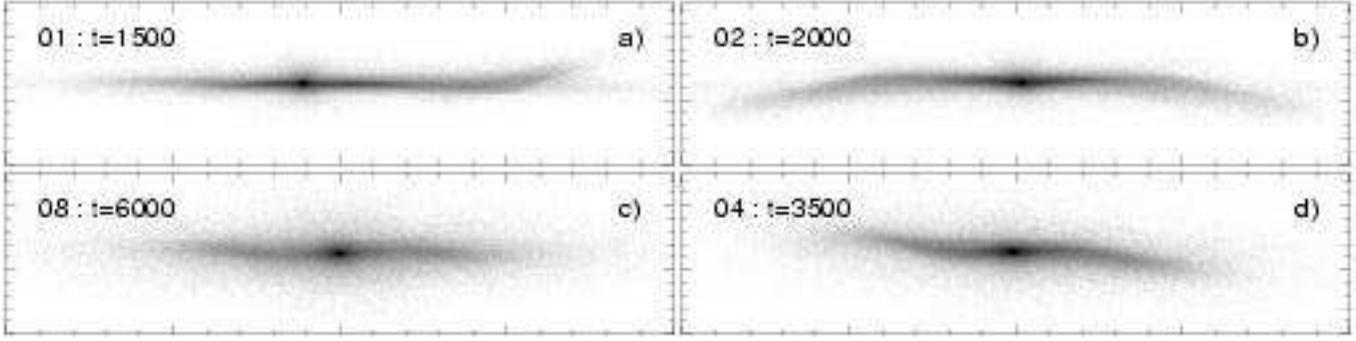}}
\caption{Edge-on projections of the models 01, 02, 08, and 04. Times are indicated at the upper left.
  The box dimensions are $100 \times 25\,\rm{kpc2}$}
\label{snapshots}
\end{figure*}
\subsection{Results}

\begin{figure*}[!]
\centering
\includegraphics[width=17.99cm]{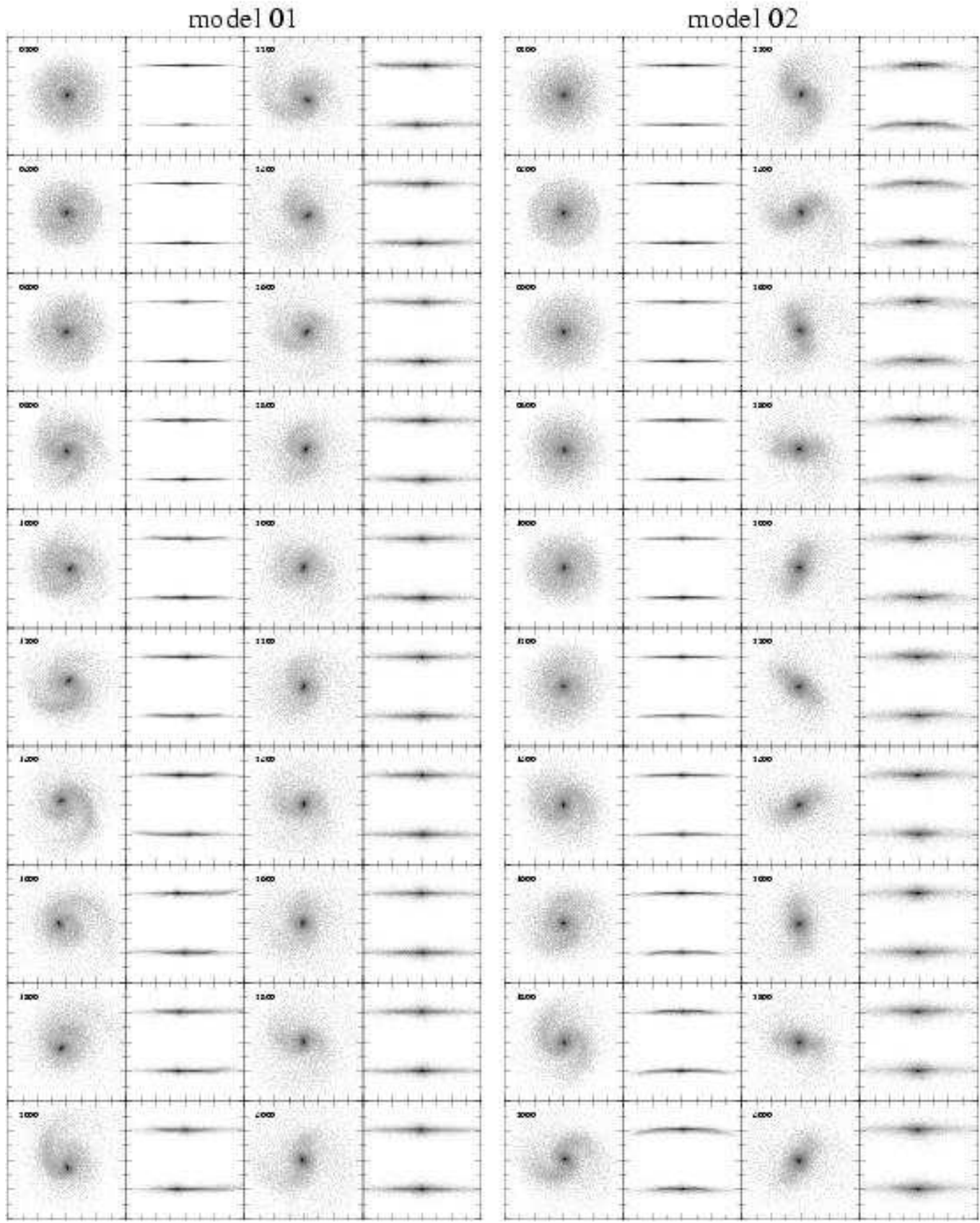}
\caption{Evolution of model 01 and 02 between $t=200$ and $t=4000$, noted in the upper left corner.
Columns 1 to 4 show the face-on and two edge-on (perpendicular) projections of model 01,
Columns 5 to 8 correspond to model 02. 
The boxes dimensions are $50\times 50\,\rm{kpc^2}$}
\label{evol12}
\end{figure*}
\begin{figure*}[!]
\centering
\includegraphics[width=17.99cm]{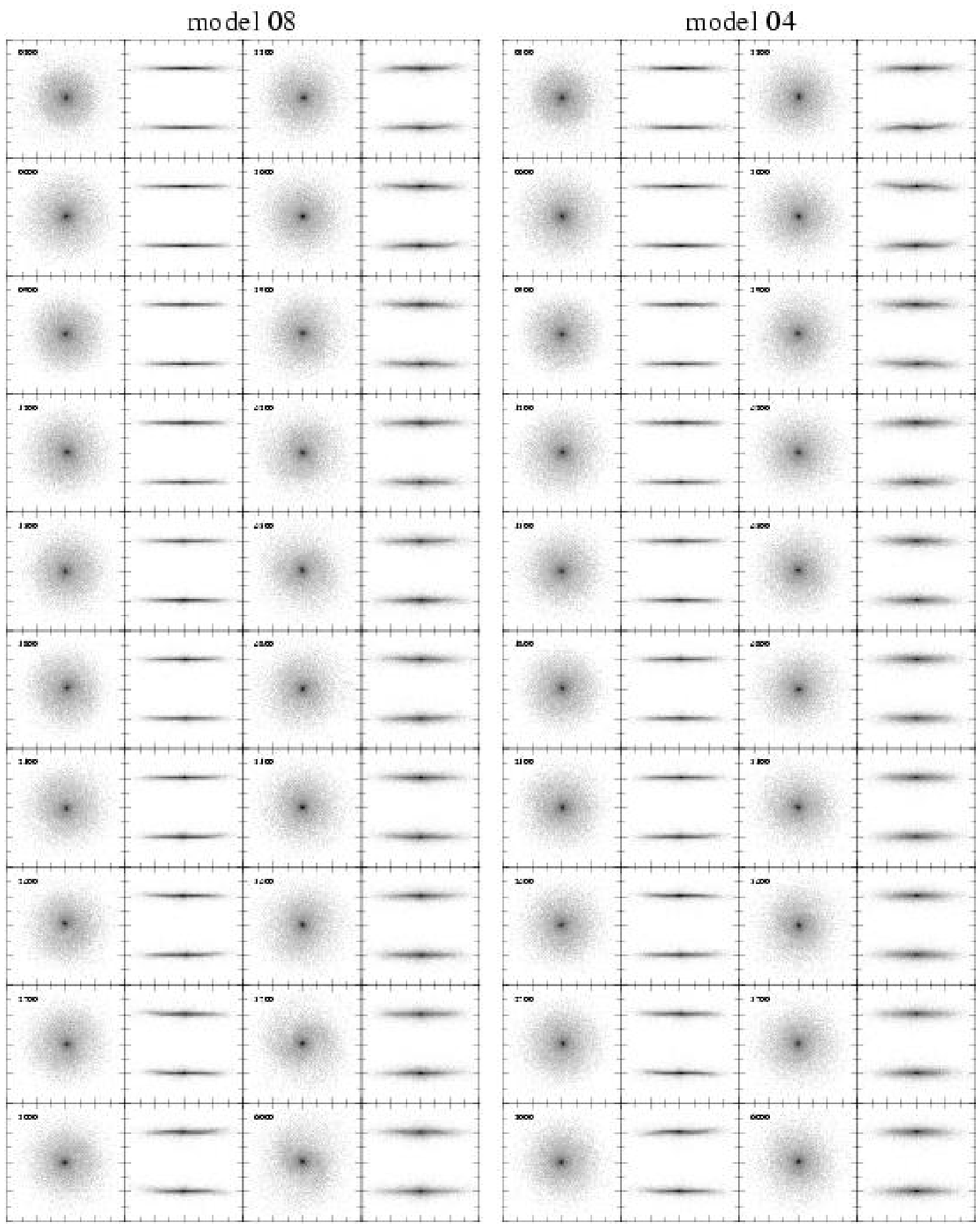}
\caption{Evolution of model 08 and 04 between $t=300$ and $t=6000$.
Columns 1 to 4 show the face-on and two edge-on (perpendicular) projections of model 08,
Columns 5 to 8 correspond to model 04. 
The boxes dimensions are $50\times 50\,\rm{kpc^2}$}
\label{evol84}
\end{figure*}

As expected, the evolution of the models depends strongly on the
thickness of the heavy disk.  In Fig.~\ref{szszr2}, from left to
right, models can be divided in four groups:

1) Model 01 has a ratio $\sigma_z/\sigma_R$ of $0.18$.  It is very
unstable. The bending instability occurs quickly and generates a
transient asymmetric warp that extends up to $z=4\,\rm{kpc}$ at
$R=35\rm{kpc}$ (see Fig.~\ref{snapshots} a). The warp is associated
with a spiral arm and its lifetime is about $1.5\,\rm{Gyr}$. The ratio
$\sigma_z/\sigma_R$ is increased by the instability above Araki's
limit, at about $0.4$. After $t=3000$, $\sigma_z/\sigma_R$ is
higher than $0.3$ and the disk remains vertically stable.
Face-on and edge-on projections of the model are displayed in 
Fig.~\ref{evol12}.

2) Models 02, 07 and 03 have still a ratio $\sigma_z/\sigma_R$ well
below the Araki limit. The bending instability occurs during the first
$2\,$Gyr. An axisymmetric bowl mode ($m=0$) grows during about
$1\,\rm{Gyr}$, before that $\sigma_z$ increases and stabilizes the
disk.  Fig.~\ref{snapshots} b) shows a spectacular U-shape warp of
model 02 at $t=2000$.  During the instability, a large bulge forms,
prolonged by a bar. The galaxies look like SBa types (see Fig.~\ref{evol12}, 
$t>2200\,\rm{Myr}$). In model 07,
the mode is slightly less symmetric, due to the lopsidedness of the
galaxy. A short-lived $m=1$ mode is present at the end of the
instability, during the bar formation.

\begin{figure}
\resizebox{\hsize}{!}{\includegraphics{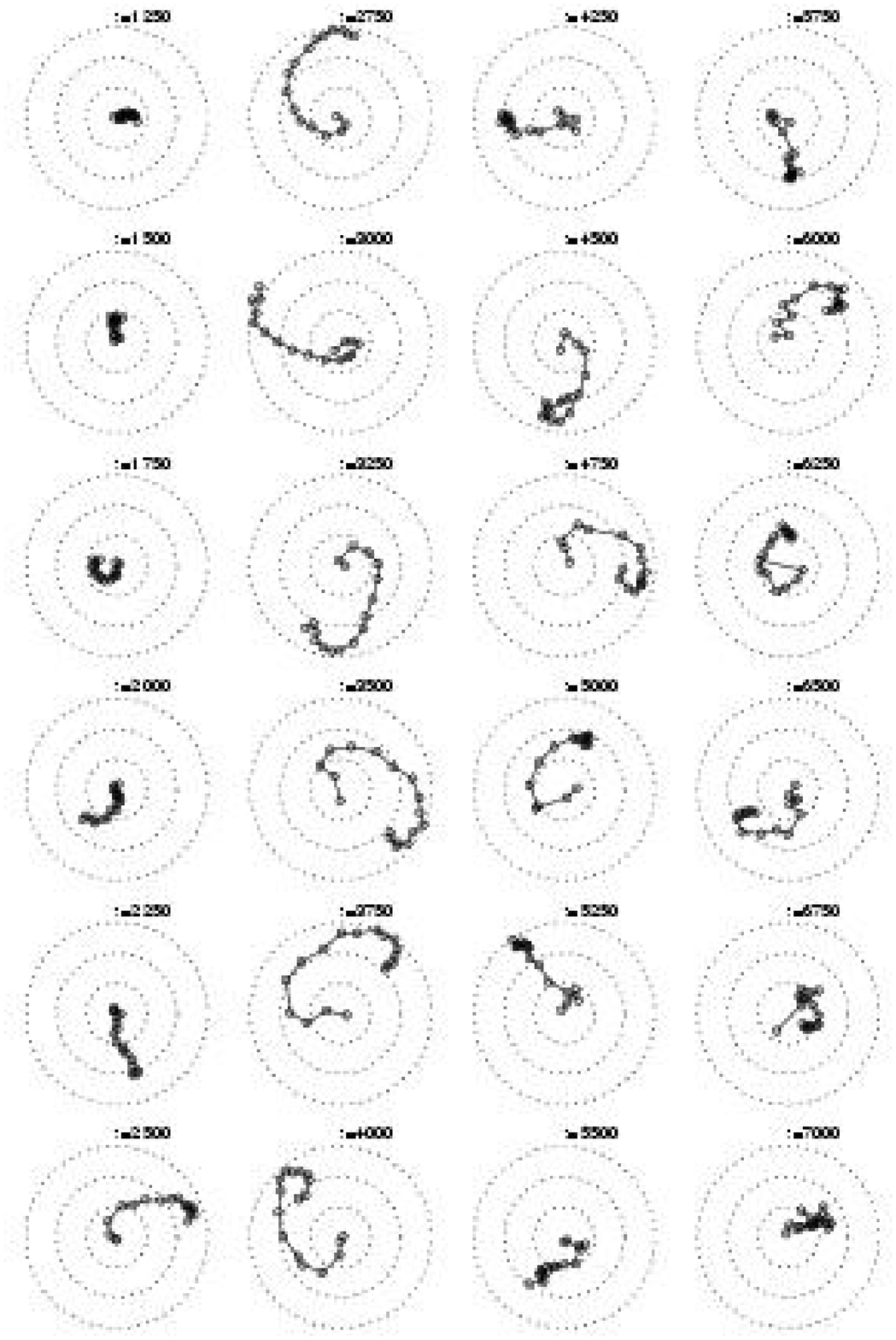}}
\caption{Evolution of the warp in model 08 from $t=1250$ to $t=7000\,\rm{Myr}$. 
  The diagrams show the polar angles of the rings (tip-LON plots).
  The three dotted circles mark $\theta=1,2,3^\circ$.}
\label{evol}
\end{figure}

3) The four models 08, 04, 09 and 05 develop S-shaped warped modes
($m=1$).  Except for model 05 which has a ratio
$\sigma_z/\sigma_R=0.3$ just above Araki's limit, all are unstable
with respect to bending.  In the case of model 08, the warp is
long-lived and lasts more than $5.5\,\rm{Gyr}$, corresponding to about
5 rotation times at $R=30\,\rm{kpc}$ (see Fig.~\ref{evol84}).  
In order to follow the
evolution of the warped disk, we have divided it in a set of 18
concentric rings from $R=0$ to $35\,\rm{kpc}$. The inclination of each
ring (warp angle) is determined by the direction of the major axis of the inertia
tensor of the matter contained in the range $[R-\Delta R,R+\Delta R]$,
where $\Delta R=1\,\rm{kpc}$ is the half width of the ring.  The
evolution of the warp is displayed in Fig.~\ref{evol} in form of
Tip-LON diagrams \citep{briggs90}.  The diagrams represent the
precession and nutation angle of each ring relative to the inner
galactic disk.  The outer dotted circle corresponds to an inclination
of $3^\circ$.  At $t=500$, the disk is flat. The warp grows from
$t=1000$ to $t=3000$ where it culminates ($z=2\,\rm{kpc}$ at
$R=35\,\rm{kpc}$). During the following $4\,\rm{Gyr}$, the amplitude
of the warp slowly decreases.  At $t=6000$ it is still clearly
observable (see Fig.~\ref{snapshots} c)).  Despite the differential
rotation of the disk, the line of nodes (LON) remains quasi straight.
However, in the outer parts, it is slightly trailing, contradicting
Briggs' third rule \citep{briggs90}\footnote{The galactic rotation
  is counterclockwise.}.  The rotation period of the LON is about
$1.1\,\rm{Gyr}$, which corresponds to the circular rotation period at
the edge of the disk ($R=35\,\rm{kpc}$).  At about $t=6500\,\rm Myr$,
the warp disappears and gives place to a large bar.  The instability
increases the vertical velocity dispersions by a factor 2 at the
maximum, around $10\,\rm{kpc}$.  At larger radius, the factor is $1.5$
or less.  Velocities in the plane are unaffected.
\begin{figure}
\resizebox{\hsize}{!}{\includegraphics{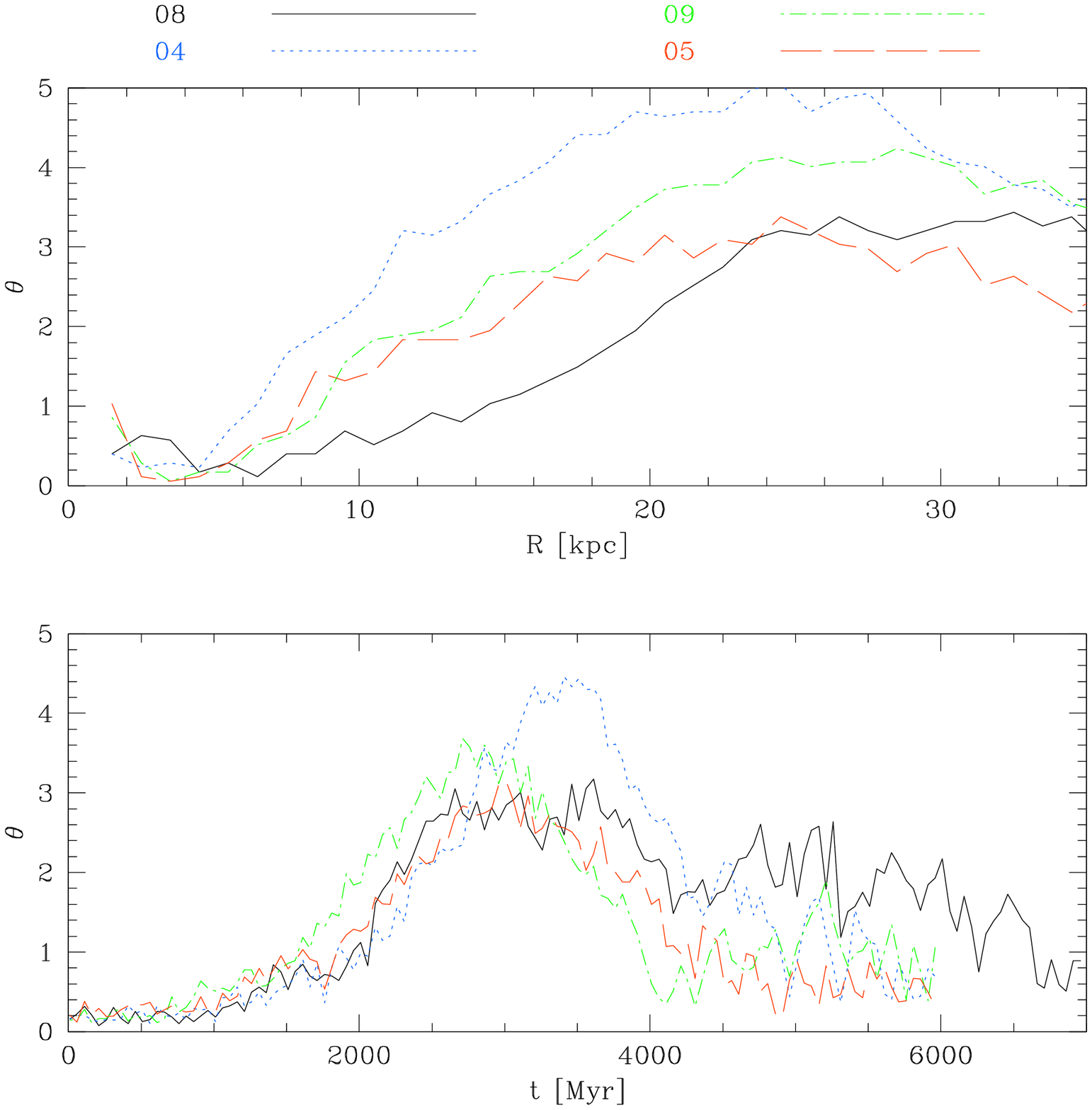}}
\caption{Top: warp angle $\theta$, as a function of radius, for the  
  models 08 ($t=3100$), 04 ($t=3500$), 09 ($t=2700$), and 05 ($t=2800$).
  Bottom: time evolution of the mean warp angle computed between
  $R=25$ and $R=35\,\rm{kpc}$.}
\label{inc}
\end{figure}
At the top of Fig.~\ref{inc}, we compare the warp angle of the four
models at the time where the warp is maximum. The warp amplitude of
model 04 is the highest. At $t=3500$ it reaches more than
$2\,\rm{kpc}$ above the plane defined by the central regions (see
Fig.~\ref{snapshots} d)). However, its lifetime is slightly shorter
than model 02 and the warp disappears after $3.5\,\rm{Gyr}$ at
$t=5000$ (see the bottom of Fig.~\ref{inc}).  The slightly thicker
model 09 and 05 also develop warps with similar amplitudes but their
lifetimes are shorter than $2.5\,\rm{Gyr}$.  In all four models,
the warp shape is characterized by an inner flat region that
extends between $5$ to $10\,\rm{kpc}$ (Fig.~\ref{inc} top).
4) Model 06 has a ratio $\sigma_z/\sigma_R$ well above $0.3$. In
this model, no vertical instability occurs and the disk remains flat all
along the simulation.

%__________________________________________________________________

\section{The influence of a dark halo}\label{halo}

%__________________________________________________________________

The previous simulations have been run in the absence of a dark halo,
which has the advantage of allowing us to identify the warp origin
uniquely to the self-gravity of the disk when Araki's instability
criterion is met.  We want now to check up to which point the disk
self-gravity leads to bending instabilities when the disk is embedded
in a conventional hot halo.

Since the bending instability is closely related to Araki's criterion,
we determine how the threshold is modified by
different flattenings and mass of a hot dark halo.
We suppose a hot halo because these are less subject to their own instabilities.
Thus, a rigid potential is a convenient approximation of the halo
role, provided the velocity dispersions $\sigma_z$, $\sigma_R$ in the
disk are calculated consistently for the new local potential.

\subsection{Models with dark halos}

The models with dark halos are based on the six first models with
$R_{f}=40\,\rm{kpc}$.  We simply add a quasi homogeneous thick mass
distribution in the disk region, however bounded in extension more
abruptly than isothermal halos because we want to transfered mass from
the heavy disk to the halo region of similar extent.  Therefore we
just adopt an inflated Plummer potential, the Miyamoto-Nagai potential
\citep{Miyamoto75} with a radial scale-length $H$ of $20\,\rm{kpc}$:
\begin{equation}
        \phi_{\rm{h}} = -\frac{GM_{\rm{h}}}{\sqrt{R^2 + \left(H-c+\sqrt{z^2 + c^2}\right)^2}}, 
\end{equation}
where $c$ is the vertical scale-height.  Here the constants in the
potential are expressed differently from the original Miyamoto-Nagai
potential in order to better express the horizontal scale length with
$H$.  The halo flattening is parametrized by $\zeta_h = c/H$. 
For $R<H$, $\zeta_h$ is a very good approximation of the ratio of minor over 
major axis of the mass iso-densities.

The mass fraction of dark matter that lies in the disk is parametrized
by the factor $f$.  We can thus write:
\begin{equation}
        M_{\rm{hd}} =  f\,M_{\rm{dark}}, \quad  M_{\rm{h}}  =  (1-f)\,M_{\rm{dark}}
\end{equation}
where $M_{\rm{dark}}=M_{\rm{hd}} + M_{\rm{h}}$.  The total mass is then:
\begin{equation}
        M_{\rm{tot}} = M_{\rm{b}} + M_{\rm{d}} + f\,M_{\rm{dark}} + (1-f)\,M_{\rm{dark}}.
\label{massfrac}
\end{equation}

\subsection{Stability criterion}

In the previous section, we have shown that the Araki limit is a good
stability criterion against bending oscillations if it is applied to
the minimum ratio $\sigma_z/\sigma_R$ along the galactic radius.

Since $\kappa^2$ and $\nu^2$ are linear with respect to the mass, we
can write (cf. Eq.~(\ref{sr})):
\begin{equation}
        \frac{\sigma_z^2}{\sigma_R^2}=
        \beta^2\,
        \frac{
          M_{\rm{b}}\,\kappa^2_{\rm{b}}
        + M_{\rm{d}}\,\kappa^2_{\rm{d}}
        + M_{\rm{dark}}[f\,\kappa^2_{\rm{hd}}  
        +          (1-f)\,\kappa^2_{\rm{h}}]            
        }{
          M_{\rm{b}}\,\nu^2_{\rm{b}}
        + M_{\rm{d}}\,\nu^2_{\rm{d}}
        + M_{\rm{dark}}[f\,\nu^2_{\rm{hd}}              
        +          (1-f)\,\nu^2_{\rm{h}}]       
        },
        \label{sr2}
\end{equation}
where the $\nu_i$ and $\kappa_i$ for each components are computed for
a unit mass component.  For the bulge, the exponential disk and the
heavy disk, those values are obtained numerically with the
Particle-Mesh potential solver.  For the halo, these frequencies are
calculated analytically.

\begin{figure*}
\subfigure[Model 01,\,$h_{z0}=0.05$]{\includegraphics[width=6.0cm,angle=-90]{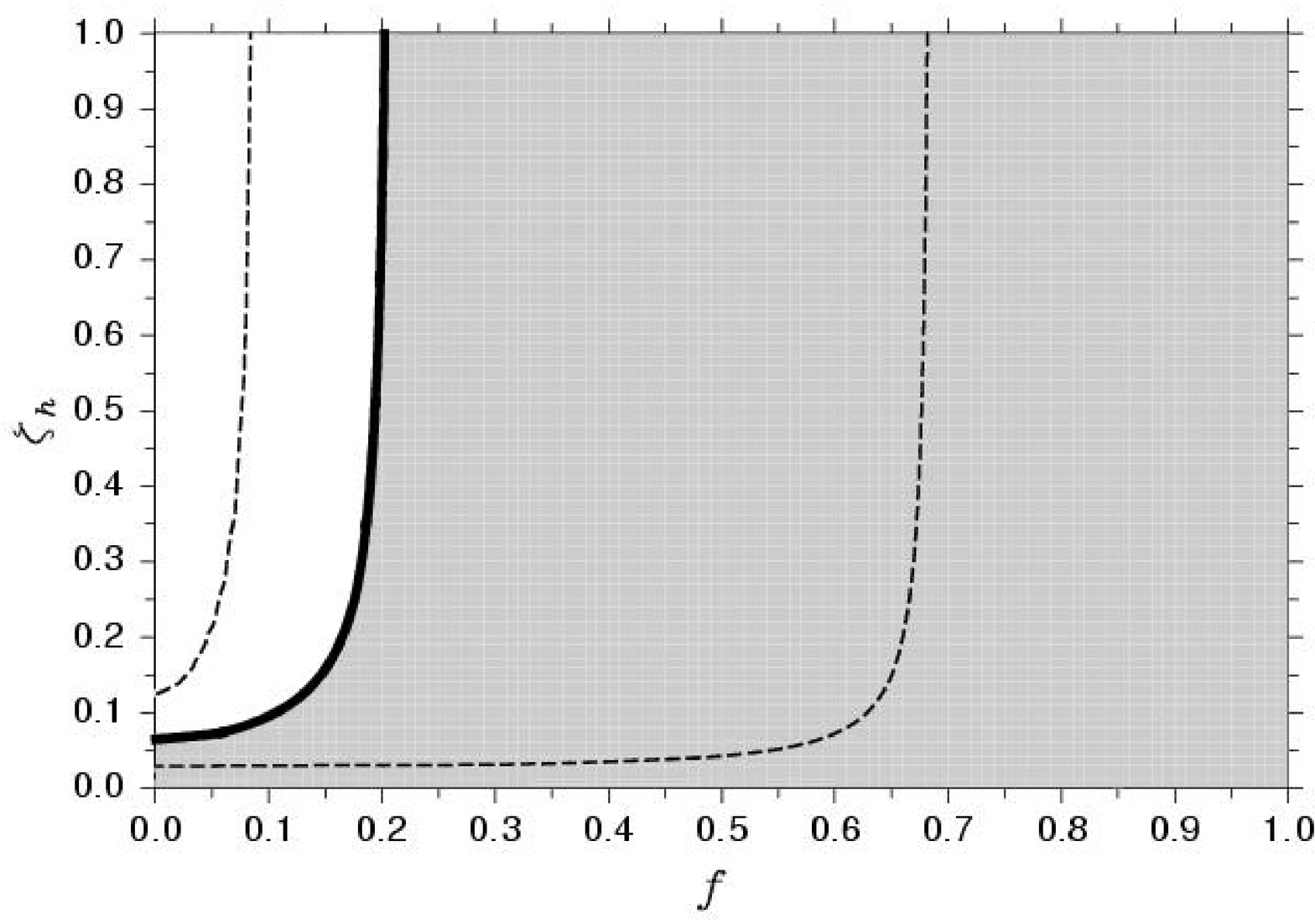}}
\subfigure[Model 02,\,$h_{z0}=0.15$]{\includegraphics[width=6.0cm,angle=-90]{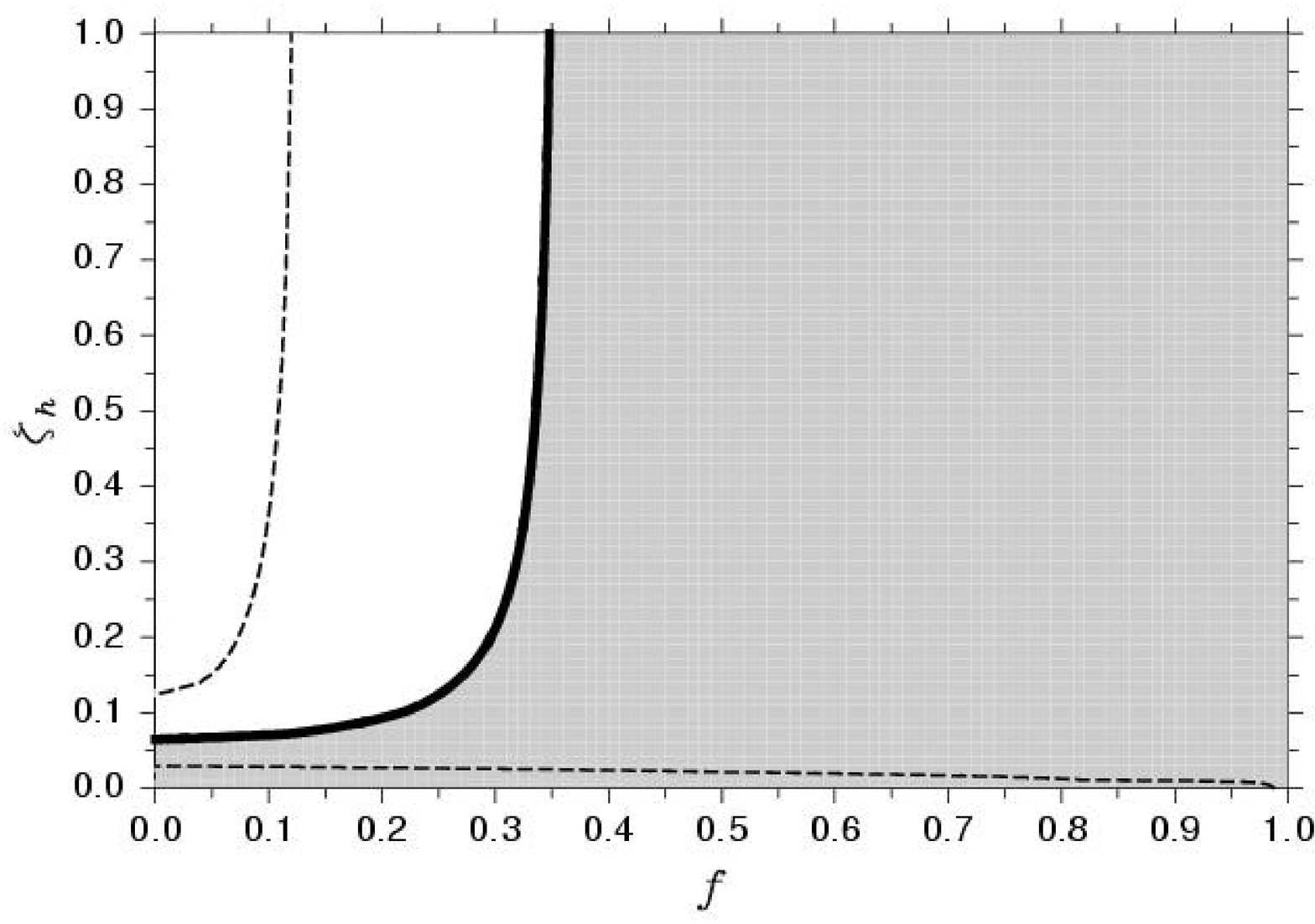}}
\subfigure[Model 03,\,$h_{z0}=0.25$]{\includegraphics[width=6.0cm,angle=-90]{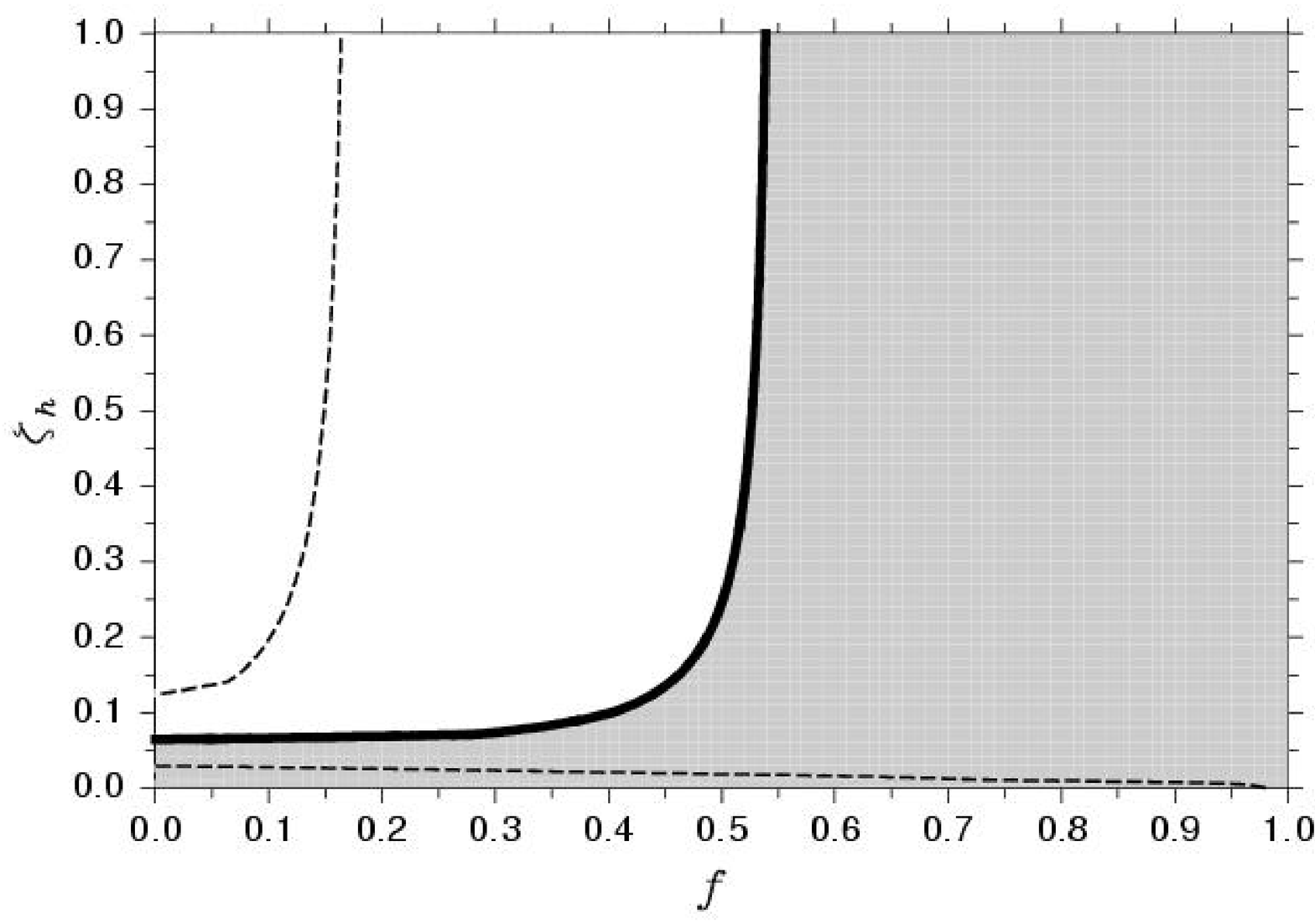}}
\subfigure[Model 04,\,$h_{z0}=0.35$]{\includegraphics[width=6.0cm,angle=-90]{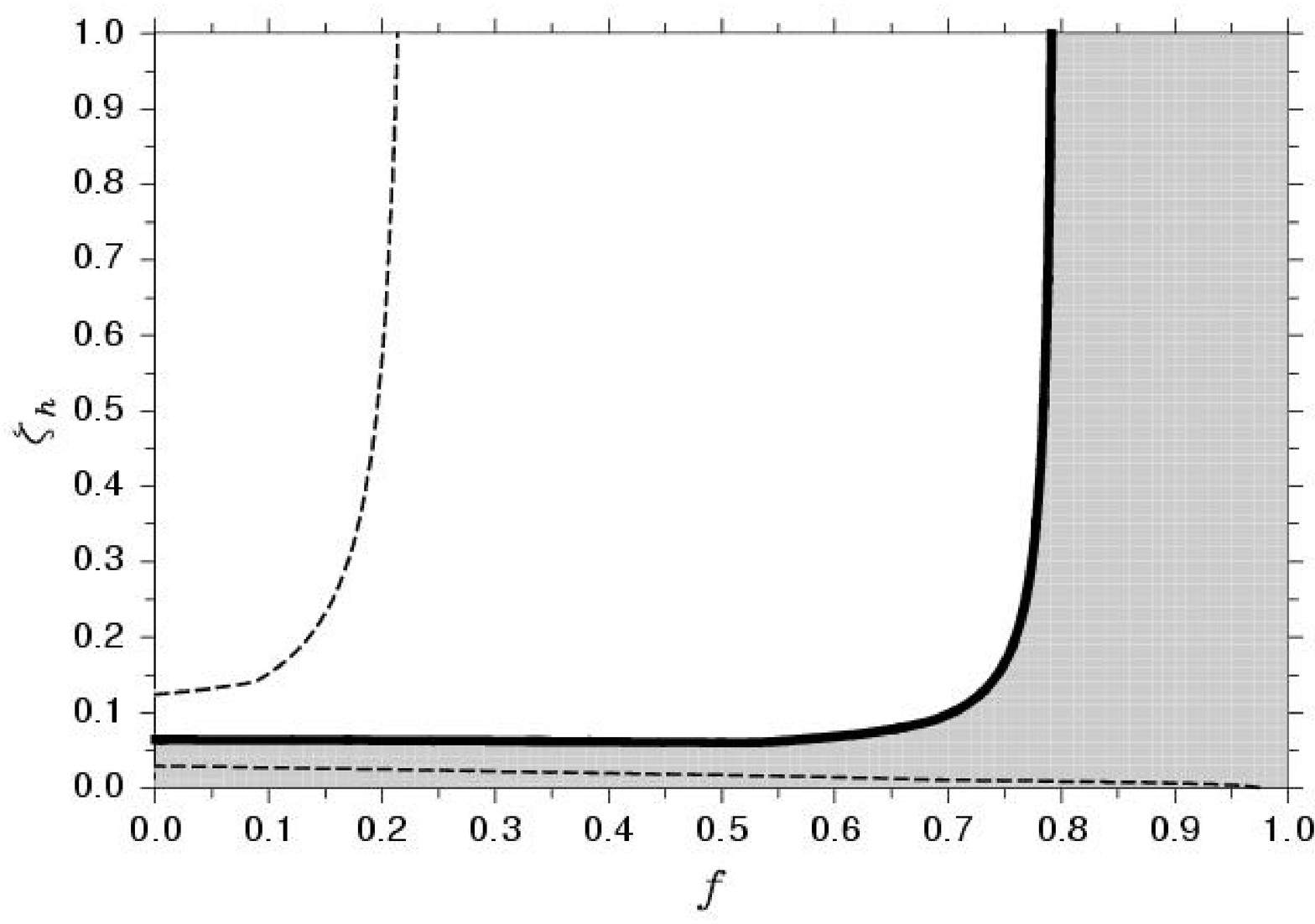}}
\subfigure[Model 05,\,$h_{z0}=0.45$]{\includegraphics[width=6.0cm,angle=-90]{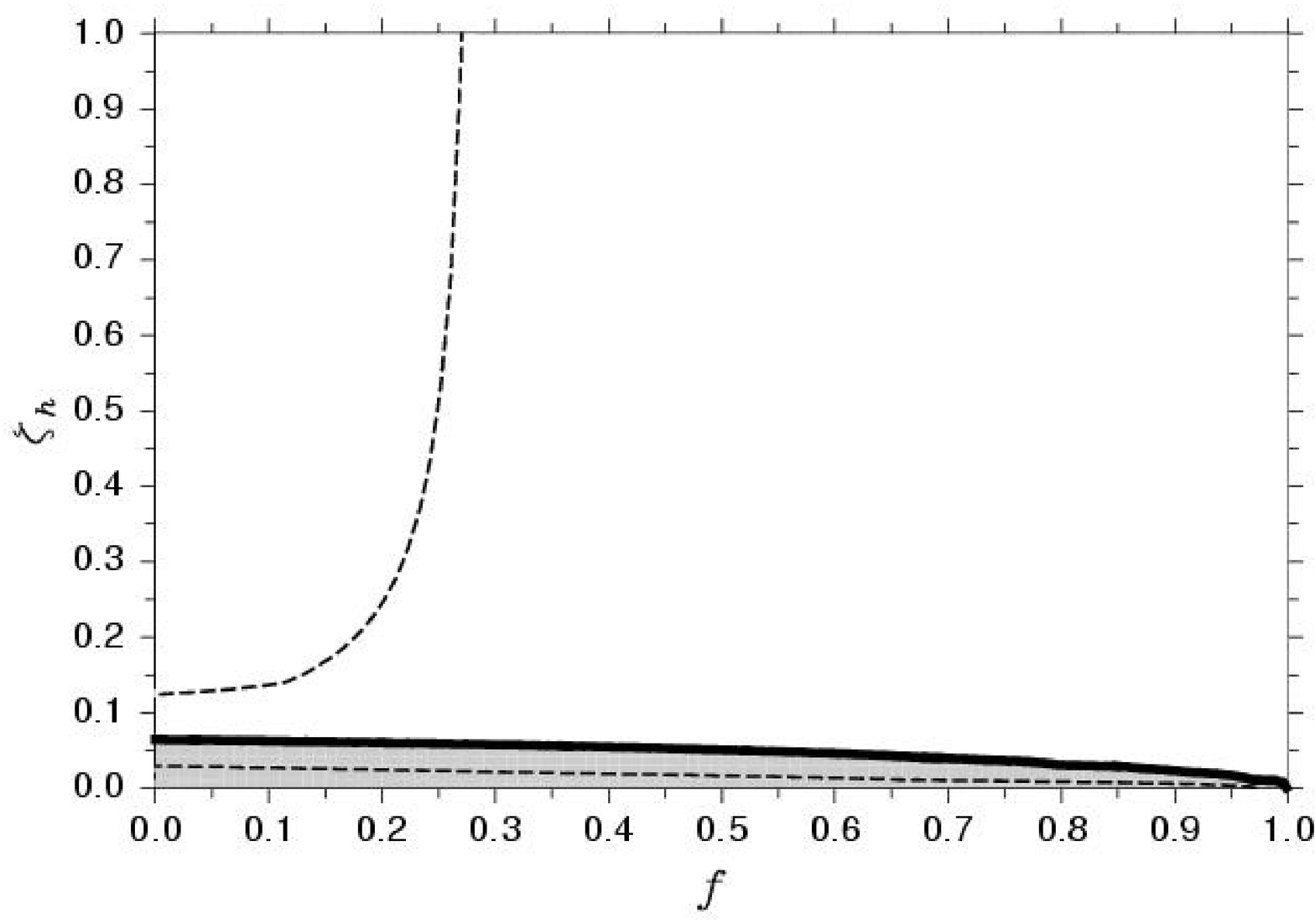}}
\subfigure[Model 06,\,$h_{z0}=0.55$]{\includegraphics[width=6.0cm,angle=-90]{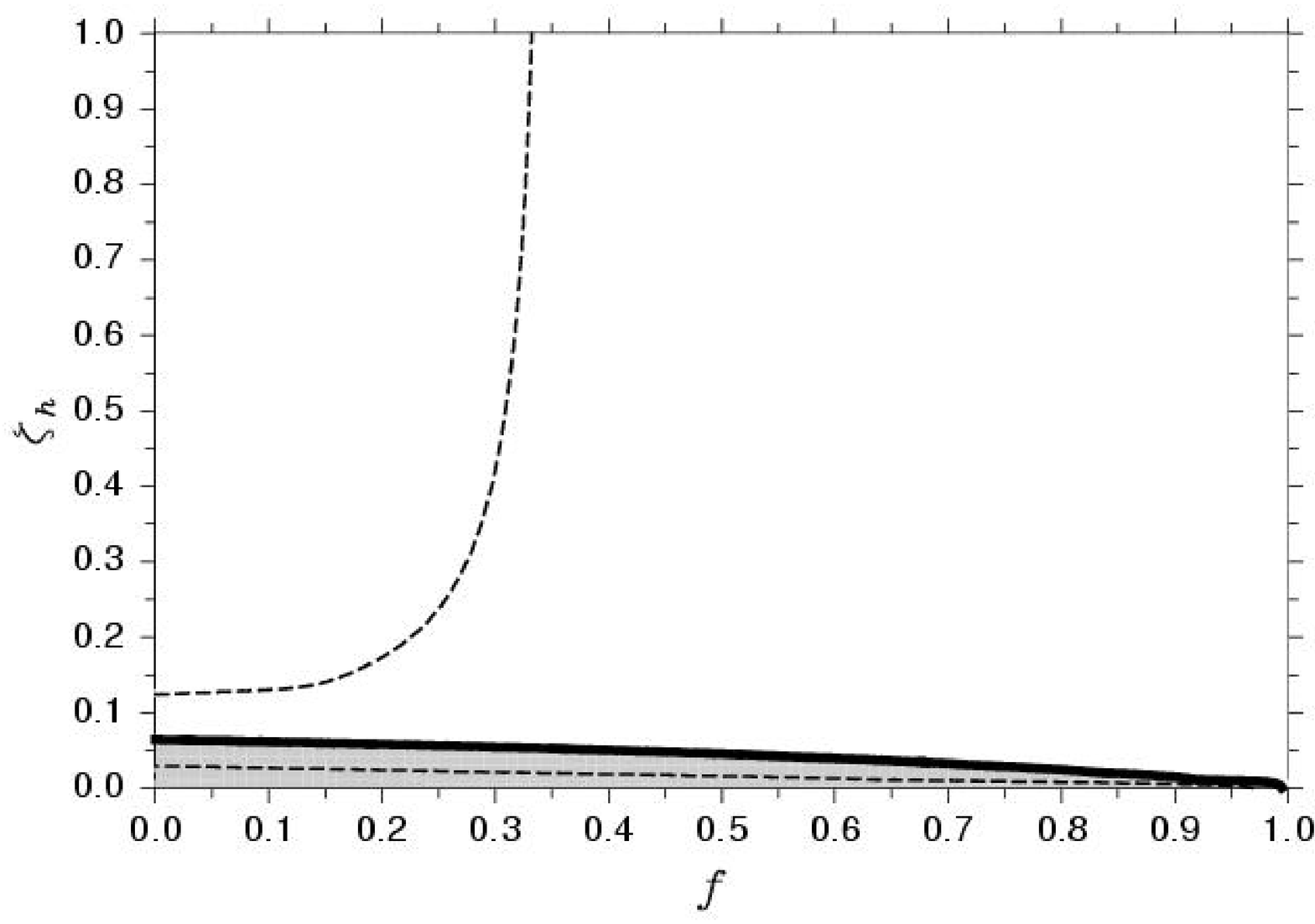}}
\caption{
  Stability of the models with dark halos against the bending
  oscillations.  The value of the minimum of
  $\frac{\sigma_R}{\sigma_z}$ along the radius, is plotted as a
  function of the fraction of dark matter in the heavy disk $f$ and of
  the halo flattening $\zeta_h$. The solid line corresponds to the
  critical value of $0.293$. The upper dashed line corresponds to the
  values of $0.4$ while bottom dashed line corresponds to $0.2$.
  Shaded regions correspond to unstable models.}
\label{araki}
\end{figure*}

For each of the 6 models, the influence of the dark matter fraction distributed in the 
disk or in the halo as well as the halo flattening is explored by varying
the parameters $f$ and $\zeta_h$ in the range $0$ to $1$. The results are
plotted in Fig.~\ref{araki}.
The Araki limit of $0.293$ is traced with the solid bold line. The models
that may be subject to bending instabilities
($\frac{\sigma_R}{\sigma_z} < 0.293$) are located at the bottom right
of this line. On the contrary, stable models
($\frac{\sigma_R}{\sigma_z}>0.293$) are located at the upper left.

In the absence of the halo ($f=1$), results corresponds to the
previous Section. Models 01 to 04 are unstable and model 06 is stable.
Model 05 is slightly above the Araki limit. However, as already mentioned,
it develops a weak instability.

When the heavy disk is replaced by a spherical halo of equal mass
($f=0$), the potential is smoothed around the $z=0$ plane and the
vertical epicycle frequency $\nu$ decreases. As $\kappa$ is weakly
modified, the ratio $\frac{\sigma_R}{\sigma_z}$ increases.
Fig.~\ref{araki} shows that the model is stable as long as $\zeta_h$
is greater than $0.06$.  If $\zeta_h< 0.06$, the halo plays the role
of the disk, and allows the bending instability to develop.  It is
relevant to note that between $\zeta_h=0.5$ and $1$, the halo
flattening has little effect on the stability.  If we consider dark
halos with $\zeta_h\ge0.5$, which corresponds to more realistic halos
according to cosmological simulations \citep{dubinski91} and weak lensing studies
\citep{hoekstra04}, the maximal halo mass in the flatter disk case
($h_{z0}=0.05$) must not be heavier than $0.8\,M_{\rm{dark}}$ ($=0.58\,M_{\rm{tot}}$). 
For thicker disks, only a halo with a mass comparable to the
heavy disk or less ($M_{\rm{h}} \le 0.36\,M_{\rm{tot}}$) allows bending 
instabilities to develop.

Some of our galaxy models can be used as first order representations
of the Milky Way.  If we assume that the Milky Way warp results from a
bending instability and that the disk is most probably critical with
respect to Araki's criterion, the solid lines in Fig.~\ref{araki}
allow us to fix an approximate constraint on the halo to disk mass
ratio.  With a vertical scale-height of about $320\,\rm{pc}$ at the
solar radius ($R_{\odot}=8.5\rm{kpc}$), model 03 has a similar
thickness to the Milky Way HI \citep{burton92}.  For any dark halo
axis ratio larger than $0.3$, the marginal stability of the disk
implies a dark halo to heavy disk mass ratio around $1$
(Fig.~\ref{araki}, (c)).  With this ratio, the surface density
($|z|<700\,\rm{pc}$) of each component in units of $\rm M_{\odot}/pc^2$
are : $\Sigma_{\rm{b}}=0.4$, $\Sigma_{\rm{d}}=36.7$,
$\Sigma_{\rm{hd}}=48.0$, $\Sigma_{\rm{h}}=3.7$, for a total surface
density of $\Sigma_{\rm{tot}}=88.8$.  The corresponding rotation
velocity at the solar radius is $205\,\rm km\,s^{-1}$.

%__________________________________________________________________

\section{Discussion}

%__________________________________________________________________

In this work, the vertical stability of a galaxy model where a
substantial fraction of the gravitational matter is contained in the
disk has been explored, leading to several aspects discussed below.

\subsection{Bending instability vs.\ disk thickness}

The strength of the investigated scenario is to require only intrinsic
gravitational dynamics, therefore the reported effects are immediately
produced whenever the stability threshold is reached, independently of
spurious external torques produced by strong accretion events, galaxy
interactions, a misaligned dark halo, or magnetic fields.  In all the
examined cases a sufficiently thin and massive disk produces a bending
instability leading to a warp.

Our results were already expected by \citet{merritt94} and
\citet{sellwood94} in studies of counter-rotating disks, that $m=0$
and $m=1$ bending modes survive in disks without counter-rotation.
Extremely thin disks with $h_{z0}<150\,\rm pc$ (potential flattening
$\zeta<0.648$) are strongly unstable and generate asymmetric transient
warps, while thick disks with $h_{z0}\ge 550\,\rm pc$ ($\zeta \ge
0.661$, $\sigma_z/\sigma_R \ge 0.3$) are completely stable.  In between,
$m=0$ (U-shaped warp) and $m=1$ (S-shaped warp) modes are
excited. $m=0$ modes grow for about $1\,\rm{Gyr}$ and disappear
when the ratio $\sigma_z/\sigma_R$ exceeds $0.3$.  S-shaped warps
appear for thicker disks ($h_{z0}=250-450 \,\rm pc$,
$\zeta=0.652-0.658$).  They are characterized by a flat central disk
and persistent straight LON, slightly trailing at large radius. The
LON rotates at the circular velocity period of the disk edge.  The
most interesting result is that in model 08, the warp is persistent
during $5.5\,\rm{Gyr}$, a substantial fraction of the galaxy life.  It
is not clear if this persistent mode may be akin to the flapping mode
reported by \citet{sellwood96} in axisymmetric disks.

In our simulations, the two extreme disks (model 01 and model 06)
differ in their density thickness by a factor of $11$. This translates
to a potential flattening variation of only $2.5$\%.  It is striking
that observed galaxies have typical HI disk thickness in the range
where bending instabilities can be triggered, because this transverse
critical state would be similar to the disk radial state, which is
also critical with respect to spiral arm formation.  

Indeed, in more realistic models taking into account the dissipational
behavior of the gas, energy dissipation reduces the gas velocity
dispersion, and gaseous disks can be expected to reach a regime
marginally unstable with respect to bending instabilities. This is
similar to the bar and spiral instabilities in the plane which heat
radially a weakly dissipative disk to a marginal stability $Q\sim 1-2$
state. Thus, we must expect that in more realistic models including
gas dissipation, bending instabilities will be repeatedly excited.

Thus, a unifying picture of galactic disks emerges as dissipative
systems maintained for a long time in a marginal stability state,
where the effects of energy dissipation are counter-balanced by
dynamical instabilities in both the radial and vertical directions.
Obviously star formation must also play an important
role in compensating energy dissipation.

\subsection{Comparison with observed warps}

This idea of connecting gas dissipation to warps is supported by
observations.  Warps are more frequent among late type galaxies
\citep{bosma91} and they are completely absent among lenticulars
characterized by gas deficiency \citep{sanchez03}.

It is also noteworthy that the present scenario predicts a predominant
occurrence of S-shaped warps, because in a slowly flattening disk due
to energy dissipation, after the Araki instability threshold is crossed
the S-shaped warps occur before the U-shaped ones.

As seen previously, only warps with modest amplitude are spontaneously
generated, with warp angle less than $5^\circ$.  Extreme cases, like
for example NGC\,4013 \citep{bottema87}, which has a $25^\circ$ warp
angle, seem to require stronger causes than an internal instability,
such as tidal interactions.  However, as the warp angle distribution
in optical surveys peaks at $3^\circ$ \citep{reshetnikov98}, this
scenario may contribute to a large fraction of the S-shaped, the
U-shaped, as well as the asymmetric warps observed in the optical.

\subsection{Relationship between warps and dark halos}

The semi-analytic analysis  in Sect.~\ref{halo} confirms our expectation that bending
instabilities can occur as long as the self-gravity of the disk is
locally dominant.  If we accept that most galactic disks are
thin and warped by bending instabilities, we obtain an additional
constraint on the mass fraction and thickness of a dark spheroidal
halo.  For extreme thin disks, such as model 02, the halo mass
fraction with respect to the \textit{total} mass (using
Fig.~\ref{araki} and Eq.~(\ref{massfrac})) cannot exceed 0.47, for
any halo flattening larger than $\sim 0.3$.  For more typical disk
thicknesses, such as in model 03, the maximum halo mass fraction drops
to 0.33, for any halo flattening larger than $\sim 0.3$.  The only way
to obtain ``classical'' halos with a mass fraction of 0.9 is to decrease
the halo flattening well below 0.1.

It is interesting to note that if now we assume that the warps are
produced by an almost marginal instability state, the above halo mass
fraction limits become equalities for any flattening above $\sim
0.3$. Therefore the warps provide a sharp constraint on the halo relative mass,
but a weak one on the halo flattening when the latter is above $\sim
0.3$.
  
These results corroborate the recent independent study of
\citet{tyurina04} where the halo mass of observed edge-on galaxies is
estimated from the thickening of the disk due to the bending
instabilities.  

\section{Conclusion}

The most important result of this study is that long-lived galactic warps
of  observed amplitudes can be produced by internal bending
instabilities, and that the dark matter to the total mass fraction
within the warp radius and included in an extended hot halo of
similar extent is limited to about a factor of 2 if warps result from a
marginal bending instability. Recent studies linking the spiral arms
and kinematic properties of disk galaxies and the Milky-Way converge
towards similar low values for dark halo masses within the HI disk
range \citep{fuchs03, kalberla03, masset03}.

Further simulations combining a heavy disk and a live dark halo
including dissipative gas or star formation feedback will
allow us to specify in more detail these first order
dynamical effects.

%__________________________________________________________________

\begin{acknowledgements}
      This work as been supported by the Swiss National Science
      Foundation.  We thank Roger Fux for interesting discussions.
\end{acknowledgements}

\end{document}